\documentclass[twocolumn,showpacs,prl]{revtex4-1}
\usepackage{epsfig}
\usepackage{graphicx}

\newcommand{\vf}{{\mbox{\boldmath$f$}}}
\newcommand{\vR}{{\mbox{\boldmath$R$}}}

\newcommand{\vk}{{\mbox{\boldmath$k$}}}
\newcommand{\vv}{{\mbox{\boldmath$v$}}}

\newcommand{\vB}{{\mbox{\boldmath$B$}}}

\newcommand{\ve}{\mbox{\boldmath$e$}}
\newcommand{\vg}{\mbox{\boldmath$g$}}

\newcommand{\vdel}{\mbox{\boldmath$\Delta$}}

\newcommand{\vsig}{\mbox{\boldmath$\sigma$}}
\newcommand{\vnab}{\mbox{\boldmath$\nabla$}}
\newcommand{\vd}{\mbox{\boldmath$d$}}

\begin{document}

\author{Z. Zheng and D.F. Agterberg}

\title{Generic nodeless Larkin Ovchinnikov states due to singlet-triplet mixing}

\begin{abstract}


Larkin-Ovchinnikov (LO) states typically have a singlet-gap that
vanishes along real-space lines. These real-space nodes lead to
Andreev midgap states which can serve as a signature of LO pairing.
We show that at these nodes, an odd-parity, spin-triplet component
is always induced, leading to a nodeless LO phase. We find the two-dimensional weak coupling, clean limit
$s$-wave phase diagram when this spin-triplet part is included. The
triplet component is large and increases the stability of the FFLO
phase. We also show that the spin-triplet contribution pushes the
midgap states away from zero energy. Finally, we show how our results can be explained phenomenologically though Lifshitz invariants. These invariants provide a simple approach to understand the role of
unconventional pairing states,
spin-orbit coupling, and inhomogeneous mixed singlet-triplet states that are not due to a FFLO instability. We discuss our results in the context of
organic superconductors.

\end{abstract}

\maketitle

There are strong reasons to suspect that the Fulde-Ferrell-Larkin-Ovchinnikov (FFLO) \cite{ful64,lar65} phases appear in the
quasi-one-dimensional(Q1D) Bechgaard salts (TMTSF)$_{2}$X\cite{shi07,yon08} and in the quasi-two-dimensional(Q2D) organics
$\kappa$-(BEDT-TTF)$_2$Cu(NCS)$_2$\cite{lor07} and $\lambda$-(BETS)$_2$GaCl$_4$\cite{tan02}. FFLO phases have also been argued to be of
importance in understanding ultracold atomic Fermi gases \cite{rad09,zwi06} and in the formation of color superconductivity in high density
quark matter \cite{cas04}. The understanding of these phases has become a relevant and topical pursuit in physics. A central result of
theoretical studies is the ubiquitous appearance of the LO phase, a striped superconducting phase in which the spin-singlet order parameter
vanishes spatially along lines\cite{mat07}. Indeed, it has been suggested that the observation of Andreev bound states localized at these nodes
would provide strong evidence for LO phase\cite{bur94,vor05}.

Here we argue that the spin-singlet LO phase is generically nodeless
due to the appearance of a spin-triplet component at the spatial
nodes of the spin-singlet component. We further show that the
triplet component is stabilized by "removing" the Andreev
bound states, that is, by pushing these states away from zero
energy.

We begin with a microscopic derivation of our main results. This derivation considers a 2D superconductor with spin-singlet $s$-wave and
spin-triplet $p$-wave pairing interactions.  This is followed by a phenomenological description that shows how Lifshitz invariants (LI) account
for the microscopic results and allow for a significant generalization to include the effects of unconventional pairing states, spin-orbit
coupling (SOC), and inhomogeneous singlet-triplet mixed states not due to an FFLO instability. While there have been prior studies of the role of $p$-wave interactions on the FFLO phase
\cite{shi07,shi00,aiz09,shi2000,fus05,bel07} and in a related phase in cold atoms \cite{sam06}, these studies have focussed on the high field
region near the normal to superconducting phase transition, where the gap is small.  Here we examine the low-field transition from a usual
superconductor to a LO phase which requires a solution of the non-linear  Eilenberger equations.

We use the Eilenberger equations as presented by Alexander\cite{ale85,bur94,vor05}. The central equation for the quasiclassical
Green's function $\hat{g}(\vR,\hat{\vk};i\epsilon_n)$ is
\begin{equation}
[i\epsilon_n\hat{\tau}_z-\hat{\Delta}-\hat{v},\hat{g}]+i\vv_f\cdot \vnab\hat{g}=0.
\end{equation}
where $\hat{\vk}$ is the direction of the Fermi momentum,
$\epsilon_n=\pi T(2n+1)$ are the Matsubara frequencies, $\vv_f$ is the Fermi velocity.
We denote the three Pauli matrices in particle-hole space by
$\tau_x$,$\tau_y$,$\tau_z$, and in spin space by
($\sigma_x$,$\sigma_y$,$\sigma_z$)$\equiv \vsig$. The Green's
function must satisfy Eilenberger's normalization condition
$\hat{g}^2=-\pi^2\hat{1}$.
 The quasiclassical Green's function in Nambu space is
\begin{equation}
\hat{g}=\left(%
\begin{array}{cc}
g+\vg\cdot\vsig & (f+\vf\cdot\vsig)i\sigma_y \\
i\sigma_y(f'+\vf'\cdot\vsig)  &  -g+\vg\cdot\vsig^*  \\
\end{array}%
\right).
\end{equation}
The Zeeman coupling with the magnetic field is given by
\begin{equation}
\hat{v}=\left(%
\begin{array}{cc}
\mu \vB\cdot\vsig & 0 \\
0  &  \mu \vB\cdot\vsig^*  \\
\end{array}%
\right),
\end{equation}
where $\mu$ is the magnetic moment of the electron. The order parameter matrix in Nambu space is
\begin{equation}
\hat{\Delta}(\vR,\hat{\vk})=\left(%
\begin{array}{cc}
0 & (\Delta+\vdel\cdot\vsig)i\sigma_y \\
i\sigma_y(\Delta^*+\vdel^*\cdot\vsig)  &  0  \\
\end{array}%
\right).
\end{equation}
The self-consistency relations are
\begin{equation}
\Delta(\vR,\hat{\vk})=N_0 \pi T\sum_n <V(\hat{\vk},\hat{\vk'}) f(\vR,\hat{\vk'};i\epsilon_n)>_{\hat{\vk'}},
\end{equation}
\begin{equation}
\vdel(\vR,\hat{\vk})=N_0 \pi T\sum_n <V(\hat{\vk},\hat{\vk'}) \vf(\vR,\hat{\vk'};i\epsilon_n)>_{\hat{\vk'}}
\end{equation}
where $V(\hat{\vk},\hat{\vk'})$ is the pairing interaction,  $N_0$ is the density of states at the
Fermi level, and  $<>_{\hat{\vk'}}$ denotes the average over the Fermi surface. To determine which phase is stable, we use the
free energy derived from the Luttinger-Ward functional by Voronstov and Sauls \cite{vor03}:
\begin{equation}
\Delta f(R)=\frac{1}{2}\int_0^1d\lambda T\sum_n N_0 \int
\frac{d^2p}{2\pi}Tr
\hat{\Delta}(\hat{g}_\lambda-\frac{1}{2}\hat{g}),
\end{equation}
$g_{\lambda}$ is an auxiliary propagator obtained from the solution to the Eilenberger equation with the physical order parameter scaled by the
dimensionless coupling parameter $0\leq\lambda\leq 1$,
\begin{equation}
[i\epsilon_n\hat{\tau}_z-\lambda\hat{\Delta}-\hat{v},\hat{g}_\lambda]+i\vv_f\cdot \vnab\hat{g}_\lambda=0.
\end{equation}
We include both singlet $s$-wave interactions and triplet $p$-wave interactions. We assume a 2D cylindrical Fermi surface and a paring interaction
 $V(\hat{\vk},\hat{\vk'})=V_s+V_t\hat{\vk}\cdot\hat{\vk'}$. The relative
strength of triplet interaction is given by the parameter $T_p=T_t/T_s$ where $T_s$ ($T_t$) are the $T_c$ for the singlet (triplet) pairing. Due to
spin rotational invariance, we will get equivalent results for the field chosen along any direction. We therefore set the field along $\hat{z}$
direction for convenience. However, we note that the magnetic field should be in the plane to ensure that vortices can be ignored. Similarly, we
also assume spatial variations along the $\hat{x}$ direction. The structure of the Eilenberger equations then ensure that there will be a
non-zero spin triplet component of the order parameter of the form $\vd=\hat{z}\sqrt{2}k_x/k_f\psi_z(x)=\hat{z}\sqrt{2}\cos\theta \psi_z(x)$.
More specifically the self-consistency relations become
\begin{equation}
\psi(\vR)=N_0 \pi T V_s \sum_n \int_0^{2\pi} \frac{d\theta}{2\pi}
f(\vR,\theta;i\epsilon_n),
\end{equation}
\begin{equation}
\psi_z(\vR)=N_0 \pi T V_t \sum_n \int_0^{2\pi} \sqrt{2} \cos(\theta)
\frac{d\theta}{2\pi} f_z(\vR,\theta;i\epsilon_n).
\end{equation}
This leads to the gap function $\psi + \sigma_z k\psi_z(x)$ that appears in the Eilenberger equations for $\hat{g}$.



\noindent {\it Phase Diagram}

In the vicinity of the transition from the normal state to the superconducting states, we set
$(\psi,\psi_z)=e^{iqx}(\tilde{\psi},\tilde{\psi}_z)$ and find the instability line $H_{c2}$ by solving the linear gap equation and optimizing
$H_{c2}$ with respect to $q$. The order parameter for a particular $q$, $\psi_q$, is a linear combination of the singlet and triplet parts, that
is $(\tilde{\psi},\tilde{\psi}_z)=(\alpha,\beta)\psi_q$. Due to parity symmetry, this solution has the same $H_{c2}$ as $\psi_{-q}$, which is
given by $(\tilde{\psi},\tilde{\psi}_z)=(\alpha,-\beta)\psi_{-q}$. As a consequence, just below $H_{c2}$, two solutions can appear: a solution
for which only one of $\psi_q$ or $\psi_{-q}$ is non-zero (known as the FF phase); or a solution for which both are non-zero and
$|\psi_q|=|\psi_{-q}|$ (known as the LO phase). To determine which of these phases appear at $H_{c2}$ requires an analysis beyond the non-linear
gap equation. Keeping up to order $|\psi|^4$ in the free energy, we find that both the FF and the LO phases appear. The FF phase takes up only a
small portion of the phase diagram. Nevertheless, this has an important physical consequence. In particular, if the FFLO phase is generated
created by a magnetic field applied in the plane, then an additional magnetic field applied along the $\hat{z}$ direction will lead to vortices.
The degeneracy of the FF and LO phases ensures that {\it there exists a stable vortex lattice of half-quantum vortices}, as opposed to the usual
Abrikosov lattice of full-quantum vortices \cite{agt08}. This half-quantum lattice will exist in a region near where these two phases are
degenerate.

We also compute the phase boundary from the uniform superconducting phase to the FFLO phase. In general, this requires a numerical solution of
the Eilenberger equations. We use an efficient and numerically stable method described by Schopohl\cite{sch95} in which the Eilenberger
equations are transformed to Riccati equations. The transition from the the uniform superconducting phase to the FFLO phases is found by
computing the free energy of these two phases. Fig. 1 shows the self-consistent order parameter at the transition from the uniform
superconducting state to FFLO state for $T_p=0.5$ and $T/T_s=0.2$. The spin-singlet order parameter is qualitatively similar to previous results
on the LO phase\cite{bur94}. However, the spin-triplet order parameter is maximum where the spin singlet order parameter vanishes, removing the
spatial line nodes usually predicted in the LO phase. Furthermore, we find that if the spin-singlet order parameter is chosen real, then the
spin-triplet order parameter is imaginary.  Both the phase and and the positions of the maxima of the spin-triplet order parameter are a natural
consequence of the phenomenological arguments presented later. The complete H-T phase diagrams are presented in Fig. 2 for $T_p=0.0$ and
$T_p=0.5$.

\begin{figure}
\epsfxsize=3.0 in \center{\epsfbox{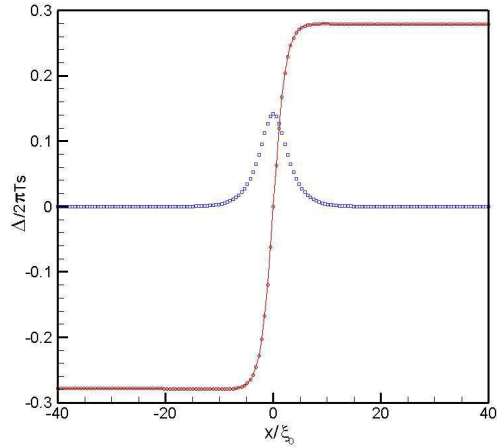}}

\caption{Singlet (circles-solid) and triplet (squares) 
order parameters at $T=0.2T_s$ for 2D FFLO superconductors
with $T_p=0.5$ and $\xi_0=v_f/(2\pi T_c)$.} \label{fig1}

\end{figure}

\begin{figure}

    \begin{tabular}{cc}
      \resizebox{80mm}{!}{\includegraphics{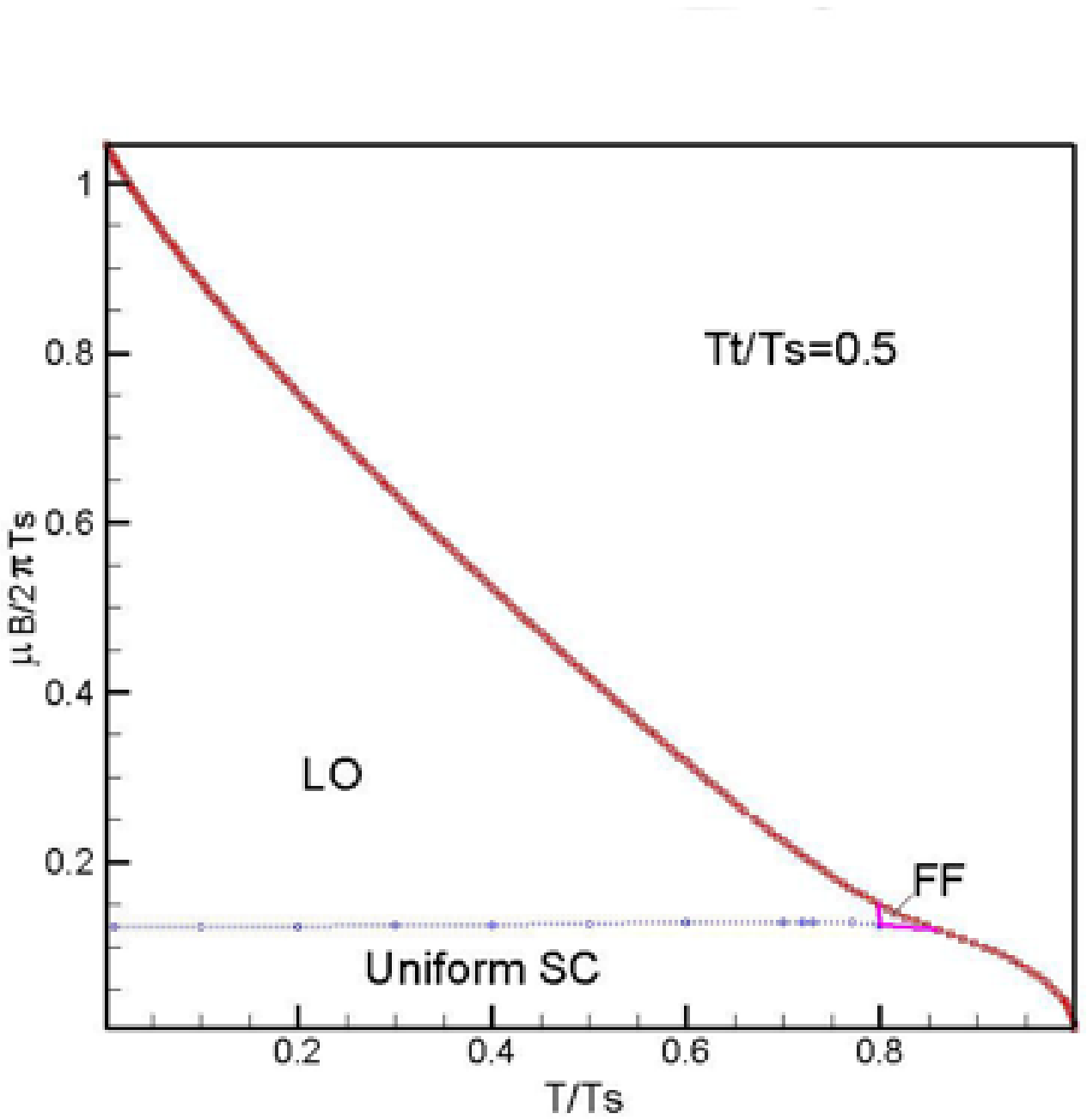}} \\
      \resizebox{80mm}{!}{\includegraphics{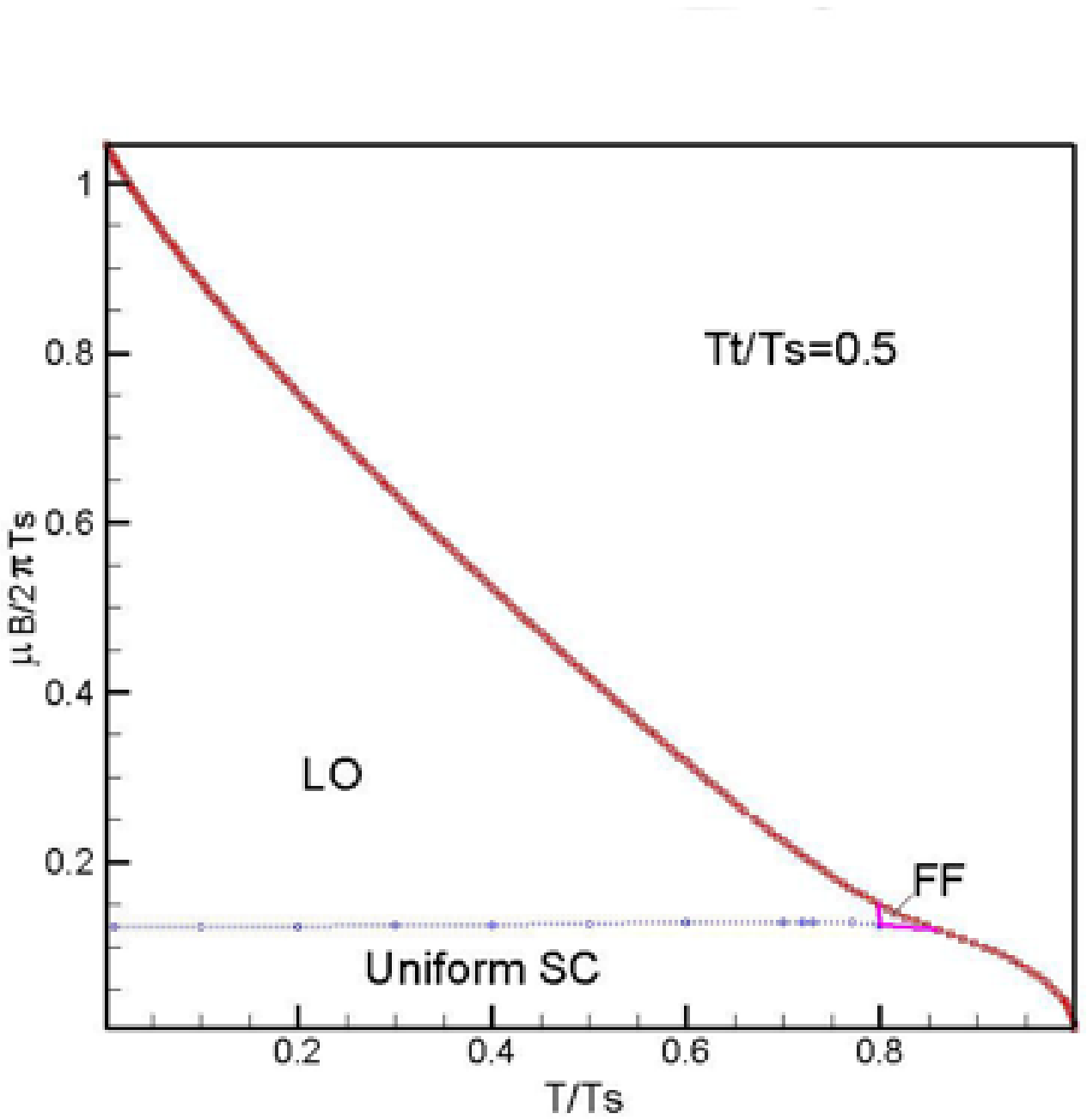}} \\

    \end{tabular}

\caption{FFLO phase diagrams for $T_p=T_t/T_s=0$ and $T_p=T_t/T_s=0.5$. At low fields, the uniform superconducting to LO phase transition
is second order (circles-dot). When $T_p=T_t/T_s=0.5$, a FF phase appears in a small region of the phase diagram (solid lines).} \label{fig2}

\end{figure}

\noindent {\it Quasiparticle Properties}

Previous studies of the LO phase have found midgap Andreev states associated with sign change of the spin-singlet order parameter
\cite{bur94,vor05}. Given the removal of the gap through the appearance of a spin-triplet order parameter, we compute the single particle
density of states to see what happens to these midgap states. The local quasiparticle density of states (LDOS) at point $\vR$ with spin
direction $\ve$ can be calculated from
\begin{equation}
N_{\ve}(\vR;\epsilon)=-<\frac{1}{\pi}Im(g(\vR,{\hat{\vk}};\epsilon)+\ve\cdot\vg(\vR,{\hat{\vk}};\epsilon))>_{\hat{\vk}}
\end{equation}
where $i\epsilon_n \rightarrow\epsilon+i0^+$. In Fig. 3, we show the LDOS at the nodes of the spin-singlet order parameter for spin-up
excitations. These results compare the solutions for $T_p=0$ and $T_p=0.5$.  The LDOS for spin-down electrons can be found by reflecting LDOS
for spin-up electrons through zero energy. When $T_p=0$, there exist Andreev bound states with energies pinned to the middle of the gap. This
agrees with previous studies\cite{bur94,vor05}. Once the spin-triplet part becomes non-zero, these states are shifted away from zero energy. This
shifting of these provides a microscopic mechanism through which the spin-triplet order parameter is energetically
stabilized. We note that a similar Andreev bound state removal mechanism has been proposed to explain the occasional appearance of spin density wave (SDW) order at the spin-singlet nodes\cite{yan09,kai09}. An important difference with our results is that the spin-triplet order we find is required to appear by
symmetry while the SDW order is not.

One physical property associated with the Andreev midgap states is the
appearance of an increased ferromagnetic magnetization at the nodes of the
spin-singlet order parameter \cite{bur94}. To investigate the role of the
spin-triplet order parameter on this, we calculate the magnetization and find that the spatial peak of
magnetization and the total magnetization both decrease due to the
shift of the Andreev states to higher energy.

\begin{figure}
 \epsfxsize=3.0 in \center{\epsfbox{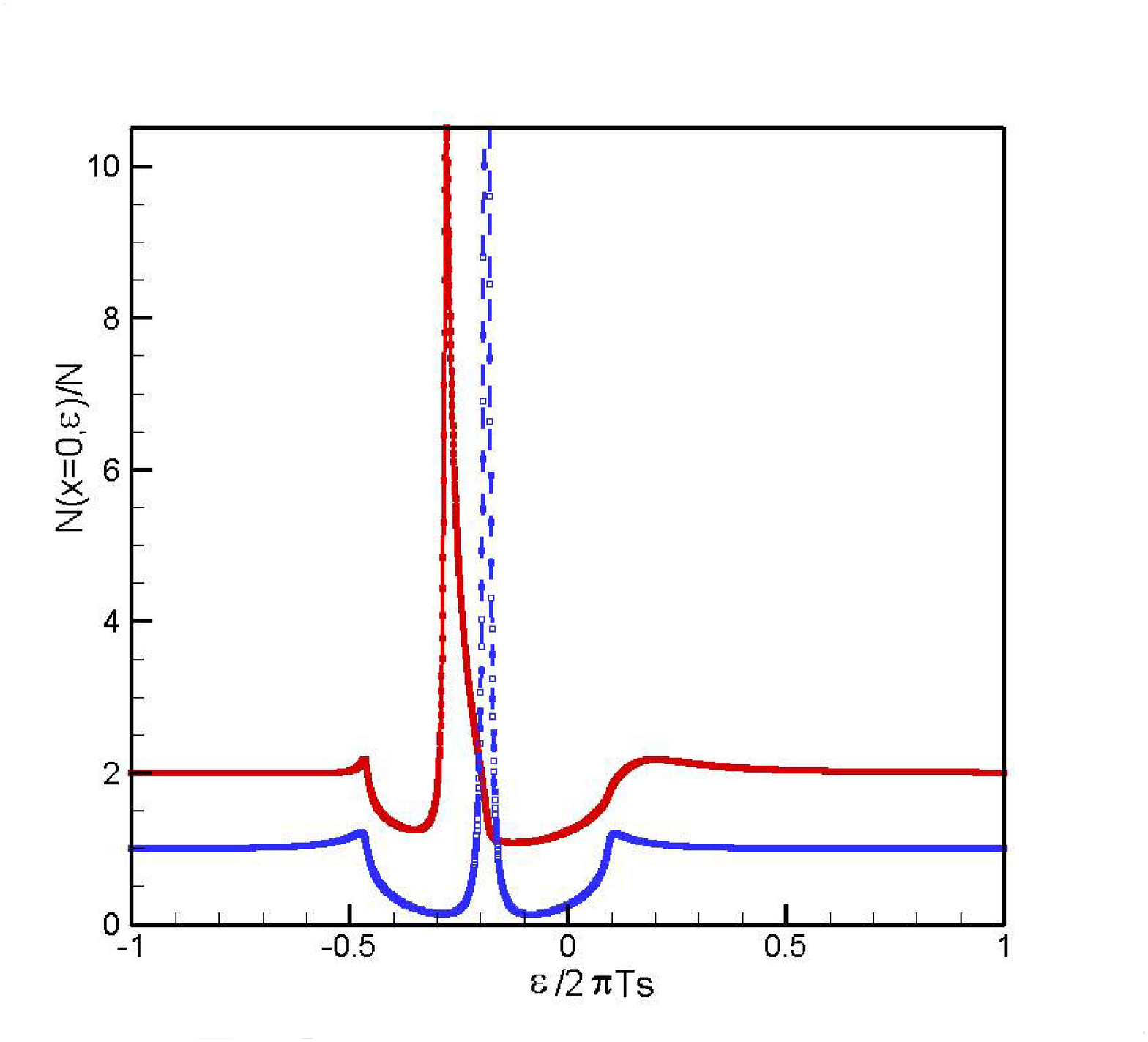}}

\caption{Spin-dependent local density of states (LDOS) at nodes of spin-singlet order parameter for $T_p=0$ (dot-dashed) and $T_P=0.5$ (solid
lines, the $y$ axis has been offset by 1 for clarity).} \label{fig3}
\end{figure}

\noindent {\it Phenomenological theory: Lifshitz invariants}

We now turn to a phenomenological description of the above microscopic results. This phenomenological theory shows that the appearance of a
spin-triplet component is generic and not specific to the microscopic details. The key point is that the admixture of spin-singlet and
spin-triplet order parameters is due to the existence of Lifshitz invariants in the Ginzburg Landau free energy (such invariants where first
discussed by Mineev and Samokhin \cite{min94}). In particular, if the spin triplet order parameter has the form $\vd(\vk,\vR)=\sum_{i,j}
A_{i,j}(\vR) \hat{x}_i k_j$ and if $\psi_s(\vR)$ describes the $s$-wave pairing, then symmetry allows the following LI (note that a similar LI
has been found in the context of cold atoms \cite{sam06})
\begin{equation}
\sum_{i,j} H_i [A_{i,j}(i\nabla_j \psi_s)^*+A_{i,j}^*(i\nabla_j \psi_s)].
\end{equation}
If, for example, the spin-singlet order parameter is given by $\psi(\vR)=\psi_0 \cos(q R_j)$, then this term implies $A_{l,j}(\vR)=\psi_{t0}
iH_l q \sin(q R_j)$ with $\psi_{t0}\ne0$. This LI ensures that a spin-triplet component is {\it always induced}. This captures some of the main
results found in the microscopic theory: the triplet order parameter is largest where the spin-singlet order parameter vanishes; and the
relative phase between the spin-singlet and spin-triplet order parameters is $\pi/2$. The LI can also be generalized to unconventional spin-singlet
order parameters and the role of SOC. For example, if the spin-singlet pairing is
$d_{x^2-y^2}$, then the following Lifshitz invariant exists
\begin{equation}
\begin{array}{cc}
\sum_i H_i[A_{i,x}(i\nabla_x\psi_d)^*+A_{i,x}^*(i\nabla_x\psi_d) \\
-A_{i,y}(i\nabla_y\psi_d)^*-A_{i,y}^*(i\nabla_y \psi_d)]
\end{array}
\end{equation}
This implies $f$-wave spin-triplet pairing appears and once again, the magnitude of the $f$-wave component is largest where the $d$-wave
component vanishes. This has been argued to be relevant in the organic (TMTSF)$_{2}$X\cite{aiz09,nic05}.  Furthermore, for example, in a
tetragonal material with spin-singlet $s$-wave order  $\psi_s$, spin-orbit interactions allow the following LI
\begin{equation}
\eta \sum_{j}[\psi_j(i\nabla_j\psi_s)^*-\psi_{j}^*(i\nabla_j \psi_s)] \label{lif2}
\end{equation}
where $\psi_i$ is defined through $\vd(\vk,\vR)=(\psi_y(\vR)k_x-\psi_x(\vR)k_y)\hat{z}$. In this case, the triplet component will have the same
phase as the s-wave component (as opposed to the $\pi/2$ phase shift for the field induced LI).

The existence of the LI plays another role not tied to the LO phase. In particular, it has been argued that a singlet to triplet phase
transition may occur in (TMTSF)$_{2}$X superconductors without the existence of a FFLO phase\cite{shi07,aiz09}.
 Such a transition is typically first order. The LI terms will transform this first order transition into a pair of second order transitions between which
 lies an inhomogeneous singlet-triplet mixed phase.
To understand this, consider adding the following simplified free energy to the LI in Eq.~\ref{lif2}
\begin{eqnarray}
f=&&\alpha_s|\psi_s|^2 +\alpha_p(|\psi_x|^2+|\psi_y|^2) +\beta_s|\psi_s|^4\nonumber \\  && +\beta_p(|\psi_x|^2+|\psi_y|^2)^2 \nonumber \\ &&+
\kappa_s |\nabla \psi_s|^2 +\kappa_p (|\nabla \psi_x|^2+|\nabla \psi_y|^2).
\end{eqnarray}
For this free energy, without the LI, the singlet to triplet phase transition is first order. Near the normal to
superconducting phase boundary, where it is sufficient to consider the quadratic free energy,
the singlet to triplet transition will occur when $\alpha_s=\alpha_t$. Close to this point, when the LI is included, the quadratic free energy is always minimized by introducing a inhomogeneous state where
both
$\psi_s\propto e^{iqr}$ and $\psi_i\propto e^{iqr}$.  This solution intervenes between the pure singlet and triplet states and the transition into this inhomogeneous phase is second
order from both the pure singlet and pure triplet phases. This indicates that even if there is no FFLO phase in (TMTSF)$_{2}$X, a closely related inhomogeneous singlet-triplet phase is likely to appear.

In conclusion, we present microscopic arguments that show that the spatial line nodes of spin-singlet LO phases are removed by the appearance of
a spin-triplet components. We show that this can be understood phenomenologically through the existence of Lifshitz invariants in the free energy which also ensure that the
spin-triplet component always appears in a spin-singlet FFLO phase. This or related inhomogeneous singlet-triplet mixed states
are likely to exist in the organic superconductors (TMTSF)$_2$X. This
work is supported by NSF grant DMR-0906655.

\end{document}